\documentclass[amsmath,amssymb,showpacs,floatfix,twocolumn,prl,superscriptaddress]{revtex4}
\usepackage{graphicx}
\usepackage{dcolumn}
\usepackage{bm}

\def\iom{i\omega}
\def\t{\rm{t}}
\def\tup{\rm{t}_\uparrow}
\def\tdo{\rm{t}_\downarrow}
\def\Li6{$^6$Li}
\def\K40{$^{40}$K}
\def\u{$|U|$\,}
\def\densup{\langle n_\uparrow \rangle}
\def\densdo{\langle n_\downarrow \rangle}
\def\ncup{n_{c\uparrow}}
\def\ncdo{n_{c\downarrow}}
\def\ndup{n_{d\uparrow}}
\def\nddo{n_{d\downarrow}}
\def\vk{\mathbf{k}}
\def\vQ{\mathbf{Q}}

\begin{document}


\title{Competing superfluid and density-wave ground-states\\
of fermionic mixtures with mass imbalance in optical lattices}
\date{\today}
\author{Tung-Lam Dao}
\affiliation{Centre de Physique Th\'eorique, \'Ecole
Polytechnique, CNRS, 91128 Palaiseau Cedex, France.}
\author{Antoine Georges }
\affiliation{Centre de Physique Th\'eorique, \'Ecole
Polytechnique, CNRS, 91128 Palaiseau Cedex, France.}
\author{Massimo Capone}
\affiliation{SMC, CNR-INFM and Dipartimento di Fisica, ``Sapienza''
Universit\'a di Roma, Piazzale Aldo
Moro 2, I-00185 Roma, Italy, and ISC - CNR, Via dei Taurini 19,
I-00185, Roma, Italy}

\begin{abstract}
We study the effect of mass imbalance on the phase diagram of a
two-component fermionic mixture with attractive interactions
in optical lattices.
Using static and dynamical mean-field theories, we show that
the pure superfluid phase is stable for all couplings when the mass imbalance is
smaller than a limiting value. For larger imbalance, phase separation
between a superfluid and a
charge-density wave takes place when the coupling exceeds a critical strength.
The harmonic trap induces a spatial segregation of the two phases,
with a
rapid variation of the density at the boundary.
\end{abstract}
\pacs{71.10.Fd, 03.75.Lm, 32.80.Pj, 71.30.+h}

\maketitle

The remarkable advances in handling ultra-cold atomic gases have
given birth to the new field of ``condensed matter physics with
light and atoms''. Cold atoms in optical lattices, with tunable
and controllable parameters, have been studied in many different
contexts (for reviews, see
\cite{lattice_reviews}).
Mixtures of two-component atoms with different masses (e.g
$^{6}$Li, $^{40}$K) introduce an additional parameter, namely the
difference between the hopping amplitudes associated with each
species in the optical lattice. This may affect the stability of
the possible quantum phases or even induce new ones. Recently, a
phase diagram has been worked out in the one-dimensional (1D)
case~\cite{cazalilla_twomasses_prl_2005}.

In this article, we consider such fermionic mixtures in higher dimensions,
with an attractive on-site coupling. Using analytical and numerical techniques, we
establish a ground-state phase diagram as a function of coupling strength and
mass imbalance, in all regimes of couplings. We also consider the experimentally relevant
effect of the trap potential, which is shown to induce a spatial segregation
between superfluid and density-wave phases.

Under conditions discussed, e.g., in
Refs.~\cite{lattice_reviews,werner_cooling_2005,duan_fermionhamiltonian_prl_2005},
fermionic mixtures are described by a Hubbard model:
\begin{equation}
H=-\sum_{\langle i,j\rangle,\sigma}\,
\t_{\sigma}(c^{+}_{i\sigma}c_{j\sigma}+\textrm{h.c.})\,-\,
|U|\sum_{i}n_{i\uparrow}n_{i\downarrow}
\label{eq:hubbard}
\end{equation}
The (pseudo-) spin index $\sigma$ refers to the two different
species. Feshbach resonances between \Li6 and \K40 are currently
under investigation~\cite{innsbruck_KLiFR_2007}, and would allow
for an attractive interaction with a tunable strength, as assumed
in (\ref{eq:hubbard}). For an example of hetero-atomic resonances
in the boson-fermion case, see e.g
\cite{stan_heteroresonance_prl_2004}.
In the following, a bipartite optical lattice made of two
interpenetrating $(A,B)$ sublattices (such as a cubic lattice) is
considered. For simplicity, we consider an equal number of atoms
for each species, leaving for future work the study of imbalanced
populations.
%

\begin{figure}[tb]
\begin{center}
\includegraphics[width=6.5cm]{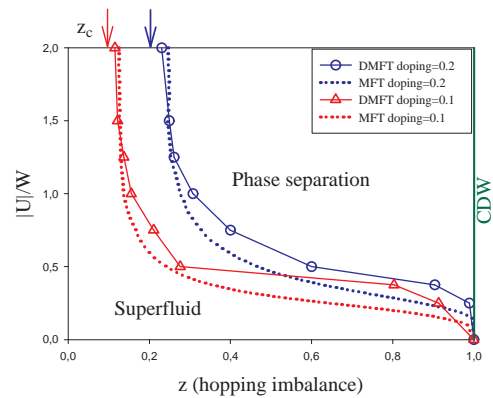}
\caption{\label{Fig:DMFT} Phase diagram of the uniform system
in the $(z,|U|)$ plane, obtained from DMFT.
Below the curves (displayed here for two `doping' levels $\delta\equiv n-1
= 0.05, 0.1$), the superfluid is stable. Above the curves, the system is
phase-separated into a half-filled CDW and a SF.
The arrows indicate the strong-coupling values obtained analytically.
The dotted lines are the weak-coupling mean-field approximation (see text).
$|U|$ is normalized to the bandwidth $W$ of
$(\varepsilon_{\vk\uparrow}+\varepsilon_{\vk\downarrow})/2$.
}
\end{center}
\end{figure}

In order to study the ground-state phase diagram of model (\ref{eq:hubbard}),
we use dynamical mean field theory at zero temperature
(DMFT)~\cite{georges_review_dmft},
together with analytical mean-field calculations for both weak
and strong coupling.
Let us anticipate the DMFT phase diagram of the uniform system,
displayed in Fig.~\ref{Fig:DMFT}. When the fermions have the same
mass, the ground-state is a superfluid (SF) for all \u. A
competing ordering exists, namely a charge density wave (CDW),
considered here in the simplest (commensurate) case in which the
charge is modulated with an alternating pattern on the $A$ and $B$
sublattices. At half-filling ($\langle n_\uparrow+n_\downarrow
\rangle=1$), it is well known that the SF and CDW states are
degenerate. This no longer applies in the `doped' system away from
half-filling: for equal masses, the SF phase is stabilized by
doping for all \u, but a large mass imbalance favors the CDW phase
over a SF state in which the Cooper pairs must be formed by
fermions with different mobilities. Hence the SF/CDW competition
becomes more interesting in the presence of mass imbalance. As
displayed on Fig.~\ref{Fig:DMFT}, we find that the uniform system
has a SF ground-state for all values of \u as long as the mass
imbalance $z\equiv (\tup-\tdo)/(\tup+\tdo)$ is smaller than a
limiting value $z_c$ (which depends on the average density). For
$z>z_c$, a (first-order) phase boundary is crossed as \u is
increased, beyond which the unform system undergoes a phase
separation (PS) between a SF and a CDW phase. As discussed later
in this paper, this implies that, in the presence of a harmonic
trap, the CDW and SF phases may both exist in different regions of
the trap.

DMFT is a quantum generalization of classical mean-field theories,
which takes the full local quantum dynamics into account, while
spatial fluctuations are neglected. It maps a lattice model onto
an effective `quantum impurity model' (a single interacting site
which hybridizes with an uncorrelated bath), subject to a
self-consistency condition~\cite{georges_review_dmft}. DMFT and
its extensions have been used to study the attractive Hubbard
model
with equal masses~\cite{negUdmft}. It is convenient to work with
Nambu's spinors $\psi^{+}=(c^{+}_{\uparrow}, c_{\downarrow})$. The
key quantity considered in DMFT is the local (on-site) Green's
function: $\hat{G}(\tau)=\langle \mathrm{T}_{\tau}
\psi_i(\tau)\psi_i^{+}(0)\rangle$ and its Fourier transform for
imaginary frequencies:
\begin{equation}
\hat{G}(\iom)=\left[%
\begin{array}{cc}
  G_{\uparrow}(\iom) & F(\iom) \\
  F^{*}(\iom) & -G_{\downarrow}(-\iom) \\
\end{array}%
\right]
\end{equation}
The superfluid order parameter is then given by $\Delta_{SF} =
\langle c_{i\uparrow}c_{i\downarrow}\rangle =
F(\tau=0)=\sum_\omega F(i\omega)$. In the CDW state, the local
Green's function takes different values ($\hat{G}_A$ and
$\hat{G}_B$) on each sublattice. The CDW order parameter is the
difference of densities on each sublattice: $\Delta_{\rm{CDW}}
=\langle n_A - n_B\rangle$.
The self-consistency conditions of DMFT relate the
(frequency-dependent) `Weiss fields'
$\hat{\mathcal{G}}_{A,B}(\iom)$ entering the effective `impurity
model' on one sublattice, to the Green's functions,
through~\cite{georges_review_dmft}:
\begin{equation}
\label{self}
\hat{\mathcal{G}}_{A(B)}^{-1}(\iom)\,=\,\iom\,\hat{1}+\hat{\mu}
-\hat{T}\,\hat{\mathbf{G}}_{B(A)}(\iom)\,\hat{T},
\end{equation}
in which $\hat{T} =
\textrm{diag}{[\t_{\uparrow},-\t_{\downarrow}]}$ and $\hat{\mu} =
\textrm{diag}{[{\mu}_{\uparrow},-{\mu}_{\downarrow}]}$ are diagonal matrices
associated with the hopping and chemical potential of each species.
As written, (\ref{self}) assumes for simplicity a semi-circular density of states,
but is easily generalized to an arbitrary lattice.
Eq.(\ref{self}) allows for the study of both SF and CDW orders,
and for their possible coexistence. The ground-state energy of the
different phases is evaluated as $\langle H\rangle=\langle
K\rangle+U\sum_{i}\langle n_{i\uparrow}n_{i\downarrow}\rangle$,
with the kinetic energy $\langle K\rangle$ in the SF and the CDW
phases reading, respectively: $\langle
K\rangle_{SF}=\beta^{-1}\sum_{\omega,\sigma}\rm{t}_\sigma^2
[G^{2}_{\sigma}(\sigma\iom)-F^{2}(\iom)]$ and $\langle
K\rangle_{\rm{CDW}}= \beta^{-1}\sum_{\omega,\sigma}
\rm{t}^{2}_{\sigma}G_{A\sigma}(\iom)G_{B\sigma}(\iom)$.

We performed DMFT calculations~\footnote{The DMFT equations
were solved using exact diagonalization~\cite{georges_review_dmft},
with $8$ energy levels in the effective bath.}
spanning the whole range of coupling $|U|$ and imbalance $z$. We
focused on the vicinity of half-filling, and found the phase
diagram of the uniform system (Fig.~\ref{Fig:DMFT}) to be
qualitatively independent of the `doping level' $\delta=\langle
n_\uparrow+n_\downarrow-1\rangle$. For small enough values
$z<z_c(\delta)$ of the mass imbalance, a pure SF solution is
stable for all $|U|$. In contrast, for $z>z_c$, the pure SF phase
is stable only for small interactions (below the line drawn in
Fig.\ref{Fig:DMFT}). Above this line (which depends on $\delta$),
the pure SF solution becomes unstable towards phase separation
between a SF and a CDW phase. (Note that we did not find a
homogeneous CDW solution out of half-filling, except at $z=1$).
This means that it is more convenient to separate the system into
a fraction $1-x$ with CDW order and $\delta=0$, and a fraction $x$
with SF order accommodating the rest of the particles.
This conclusion is reached by minimizing over $x$ the
expression $E_{\rm{PS}}(x)=(1-x) E_{\rm{CDW}}+x E_{SF}$.
We note that the SF phase is more stable than in the
1D case~\cite{cazalilla_twomasses_prl_2005} (in which nesting
favors a CDW with $\mathbf{Q}=2k_F$).

In the following, we describe analytical mean-field calculations
for both weak and strong coupling which help in understanding the
DMFT phase diagram established numerically.
We first present a {\it strong-coupling} analysis, which
holds for $|U|\gg\t_{\uparrow},\t_{\downarrow}$.
In order to analyze this limit, we find it useful to resort to a
particle-hole transformation (Table~\ref{table:ph}) and
work in the repulsive-$U$ framework. We emphasize that we are not
switching to truly repulsive interactions, but we simply exploit a
mapping.
\begin{table}[h!]
\begin{tabular}{|c|c|}
\hline
$-|U|<0$ &  $|U|>0$ \\
\hline
$c^{+}_{i\uparrow}$, $ c^{+}_{i\downarrow}$ &   $d^{+}_{i\uparrow}$, $(-1)^{i}d_{i\downarrow}$ \\
$n_{c\uparrow}$, $n_{c\downarrow}$ &  $n_{d\uparrow}$, $1-n_{d\downarrow}$ \\
$\delta\equiv n_c-1=
\langle\ncup+\ncdo\rangle-1$ & $m_d=\langle\ndup-\nddo\rangle$\\
chemical potential : $\mu_c$ & field: $h_d=\mu_c-|U|/2$ \\
$h_c$ & $\mu_d=h_c+|U|/2$\\
SF: $\langle c^+_{i\uparrow}c^+_{i\downarrow}\rangle$ &
SDW$_{xy}$: $(-1)^i\langle d^+_{i\uparrow}d_{i\downarrow}\rangle$
\\
CDW: $(-1)^i \langle \hat{n}_{ci}\rangle$ &
SDW$_{z}$: $(-1)^i \langle S^z_{di}\rangle$
\\
 \hline
\end{tabular}
\caption{Particle-hole transformation mapping the $U<0$ model
with $\densup=\densdo$ onto a half-filled $U>0$ model with
a magnetic field.}
\label{table:ph}
\end{table}
Under this mapping, our model is transformed, at large
$|U|\gg\t_{\uparrow},\t_{\downarrow}$, into an XXZ quantum spin-$1/2$
model~\cite{cazalilla_twomasses_prl_2005,duan_fermionhamiltonian_prl_2005}:
\begin{equation}
\label{XXZ}
H=J\sum_{\langle i,j\rangle}\vec{S_{i}}.\vec{S_{j}}+\gamma
J\sum_{\langle
i,j\rangle}S^{z}_{i}S^{z}_{j}-h\sum_{i}(2S^{z}_{i}-m)
\end{equation}
in which $\vec{S}\equiv\frac{1}{2}d^{+}_{\alpha}\vec{\sigma}_{\alpha\beta}d_{\beta}$,
$J=4t_{\uparrow}t_{\downarrow}/|U|$ and
$\gamma=(t_{\uparrow}-t_{\downarrow})^{2}/2t_{\uparrow}t_{\downarrow}=2z^{2}/(1-z^{2})$.
Hence, the mass imbalance turns into a spin exchange anisotropy.
The uniform magnetic field $h$ corresponds to the original
chemical potential $\mu-|U|/2$ and the magnetization to the doping $\delta$
(cf. Table.~\ref{table:ph}).
The mean-field approach~\cite{scalettar_bosehubbard_prb_1995} amounts to treat the
spin variables as classical, and minimize the energy over the
angles $\theta_A,\theta_B$ describing the orientation of the spins in the two
sublattices. The energy per site reads (with $\zeta$ the lattice connectivity
and $c_{A,B}\equiv\cos\theta_{A,B},s_{A,B}\equiv\sin\theta_{A,B}$):
\begin{equation}
\frac{E}{N}=\frac{\zeta}{8}J\,s_A s_B+
\frac{\zeta}{8}J(1+\gamma)c_A c_B
-\frac{h}{2}[c_A+c_B-2m]
\end{equation}
The phase diagram is characterized by the competition between the
$xy$ spin-density wave (SDW$_{xy}$) with order parameter
$\Delta_{xy}=\langle(-1)^{i}S_{i}^{x}\rangle$ (corresponding to SF
ordering for $U<0$), and N\'eel order (SDW$_z$),
$\Delta_{z}=\langle(-1)^{i}S_{i}^{z}\rangle$ (corresponding to
CDW).
The solution changes according to the magnetization $m$ of the
system (i.e. the doping of our physical model).
The $m$ {\it vs.} $h$ curve has a discontinuity of amplitude
$m_c=\sqrt{\gamma/(\gamma +2)}=z$.
For $m=0$ (half-filling $\delta=0$), a SDW$_z$ (CDW) state is obtained.
For $m\in[m_c,1]$, the homogeneous SDW$_{xy}$ (SF) state is stable,
while for $0<m<m_c$ phase separation takes place between the
two types of ordering. Thus, when working at fixed magnetization
(corresponding to fixed doping $\delta$), one finds a SF for
$z<z_c=m=\delta$ and phase separation for $z>z_c=\delta$.
This strong coupling value (indicated by arrows on Fig.~\ref{Fig:DMFT})
agrees very well with our DMFT results.

We now turn to the opposite weak-coupling limit. We decouple the
interaction term in the SF and the CDW channels, and determine the
regions of stability of each phase. We first consider the BCS
decoupling of the interaction, introducing the order parameter
$\Delta_{\rm{BCS}}=(|U|/N) \sum_{\vk}\langle
c^{+}_{\vk\uparrow}c^{+}_{\vk\downarrow}\rangle$ to make the
Hamiltonian quadratic. In Nambu formalism it reads:
\begin{equation}
\label{BCSHam}
H_{\rm{BCS}}=\sum_{\vk}\psi^{+}_{\vk}\left[%
\begin{array}{cc}
  \xi_{\vk\uparrow} & -\Delta_{\rm{BCS}} \\
  -\Delta_{\rm{BCS}} & -\xi_{\vk\downarrow} \\
\end{array}%
\right]\psi_{\vk}+E_{G},
\end{equation}
Here, $\tilde{\mu}_{\sigma}\equiv\mu-U n_{-\sigma}$,
$\xi_{\vk\sigma}=\varepsilon_{\vk\sigma}-\tilde{\mu}_{\sigma}$ and
$E_{G}=\sum_{\vk}\xi_{\vk\downarrow}+
N|U|n_{\uparrow}n_{\downarrow} + N\Delta^{2}_{\rm{BCS}}/|U|$. The
diagonalization of (\ref{BCSHam}) yields the Bogoliubov modes with
eigenvalues
$E^{\pm}_{\vk}=\pm(\xi_{\vk\uparrow}-\xi_{\vk\downarrow})/2+
\sqrt{(\xi_{\vk\uparrow}+\xi_{\vk\downarrow})^{2}/4+\Delta^{2}_{\rm{BCS}}}$.
Defining new variables
$\xi_{\vk}=(\xi_{\vk\uparrow}+\xi_{\vk\downarrow})/2,\
\tilde{\mu}=(\tilde{\mu}_{\uparrow}+\tilde{\mu}_{\downarrow})/2$,
the usual form of the BCS gap equation is recovered, and tells us
that the normal state is always unstable toward SF ordering.
Analogously, we can decouple the interaction in the CDW channel
defined by the order parameter
$\Delta_{\sigma}=(|U|/N)\sum_{\vk}\langle
c^{+}_{\vk+\vQ\sigma}c_{\vk\sigma}\rangle$ with
$\vQ=(\pi,\cdots,\pi)$. Introducing the spinor
$\psi^{+}_{\vk\sigma}=(c^{+}_{\vk\sigma},c^{+}_{\vk+\vQ \sigma})$,
the mean-field Hamiltonian reads:
\begin{equation}
H_{\rm{CDW}}=\sum_{\vk\in\rm{RBZ},\sigma}\psi^{+}_{\vk\sigma}\left[
\begin{array}{cc}
\varepsilon_{\vk\sigma}-\tilde{\mu}_{\sigma}   & -\Delta_{\sigma} \\
  -\Delta_{\sigma} &-\varepsilon_{\vk\sigma}-\tilde{\mu}_{\sigma}  \\
\end{array}
\right]\psi_{\vk\sigma}+E_{0}
\end{equation}
with
$E_{0}=N\Delta_{\uparrow}\Delta_{\downarrow}/|U|+N|U|n_{\uparrow}n_{\downarrow}$.
It is readily diagonalized, with eigenvalues:
$E^{\pm}_{\vk\sigma}=\pm\sqrt{\varepsilon_{\vk\sigma}^{2}+\Delta^{2}_{\sigma}}-\tilde{\mu}_{\sigma}$.
This yields two self-consistency conditions:
\begin{align}\nonumber
&\frac{1}{N}\sum_{\vk\in\textrm{RBZ}}[f(E^{+}_{\vk\sigma})+f(E^{-}_{\vk\sigma})]=n_{\sigma}\\
&\frac{\Delta_{\sigma}}{N}\sum_{\vk\in\textrm{RBZ}}\frac{f(E^{-}_{\vk})-f(E^{+}_{\vk})}
{\sqrt{\varepsilon_{\vk\sigma}^{2}+\Delta^{2}_{\sigma}}}=\frac{\Delta_{-\sigma}}{|U|}
\end{align}
\begin{figure}[tb]
\label{FigMF}
\includegraphics[width=6.cm]{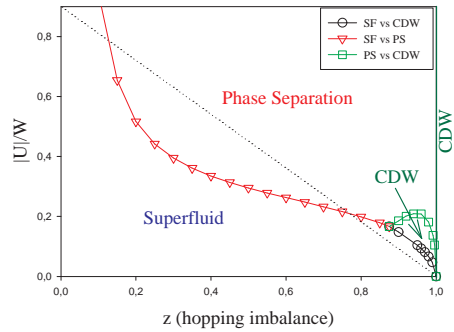}\\
\caption{Phase diagram for $\delta=0.05$ from weak-coupling
mean-field (whose validity is questionable above the dotted line)
-see text-. For simplicity, a square density of states was used
here.} \label{Fig:MFT}
\end{figure}
At a fixed value of the chemical potential, these CDW equations
have the following solutions: i) for all $|U|$ and $z$, a normal
solution with $\Delta_{\rm{CDW}}=0$, which is
unstable towards SF ii) for large enough $|U|$,
a half-filled (commensurate) CDW and iii) for large values of $z$,
close to $1$ a homogeneous CDW solution is also found with
a density different from unity ($\delta\neq 0$).

We first compare the ground-state energies of two mean-field solutions:
the homogeneous SF, and the SF/half-filled CDW phase-separated solution
obtained from a Maxwell construction. The resulting phase boundary
(Fig.~\ref{Fig:DMFT}) is seen to be qualitatively reasonable, and even
quantitatively accurate (in comparison to the numerical DMFT result)
for some intermediate range of $z$.
Indeed, the weak-coupling mean-field is justified
when $|U|\lesssim\t_{\uparrow},\t_{\downarrow}$, i.e. $|U|/W\lesssim (1-z)$.
In Fig.~\ref{Fig:MFT}, we perform a more detailed comparison of
the ground-state energies of three mean-field solutions: the
homogeneous SF, the phase separated SF/CDW, and the homogeneous
CDW with $\delta\neq 0$ (when it exists). This comparison yields a
small region of parameters, for large $z$, in which a homogeneous
CDW with a density different from one atom per site is stable.
This pure CDW pocket might be unstable to SF ordering, yielding a
candidate for a supersolid region, but we did not check this explicitly.
Anyhow, this solution is stabilized
in a region where the reliability of weak-coupling mean-field is
questionable. The lack of such a solution in DMFT may lead to the conclusion
that the CDW (or supersolid) is an artefact of weak-coupling mean-field,
but it must also be noted that the DMFT solution has a finite numerical
resolution, and that the energetic balance involved is very delicate.
Hence, we cannot reach a definitive conclusion on this issue.
Exactly for $z=1$ the ``down spin'' atoms are no longer mobile and
we have a Falicov-Kimball model, which has a pure CDW ground
state~\cite{freericks_reviewFKmodel_rmp_2003}.

\begin{figure}
  \includegraphics[width=8.6cm]{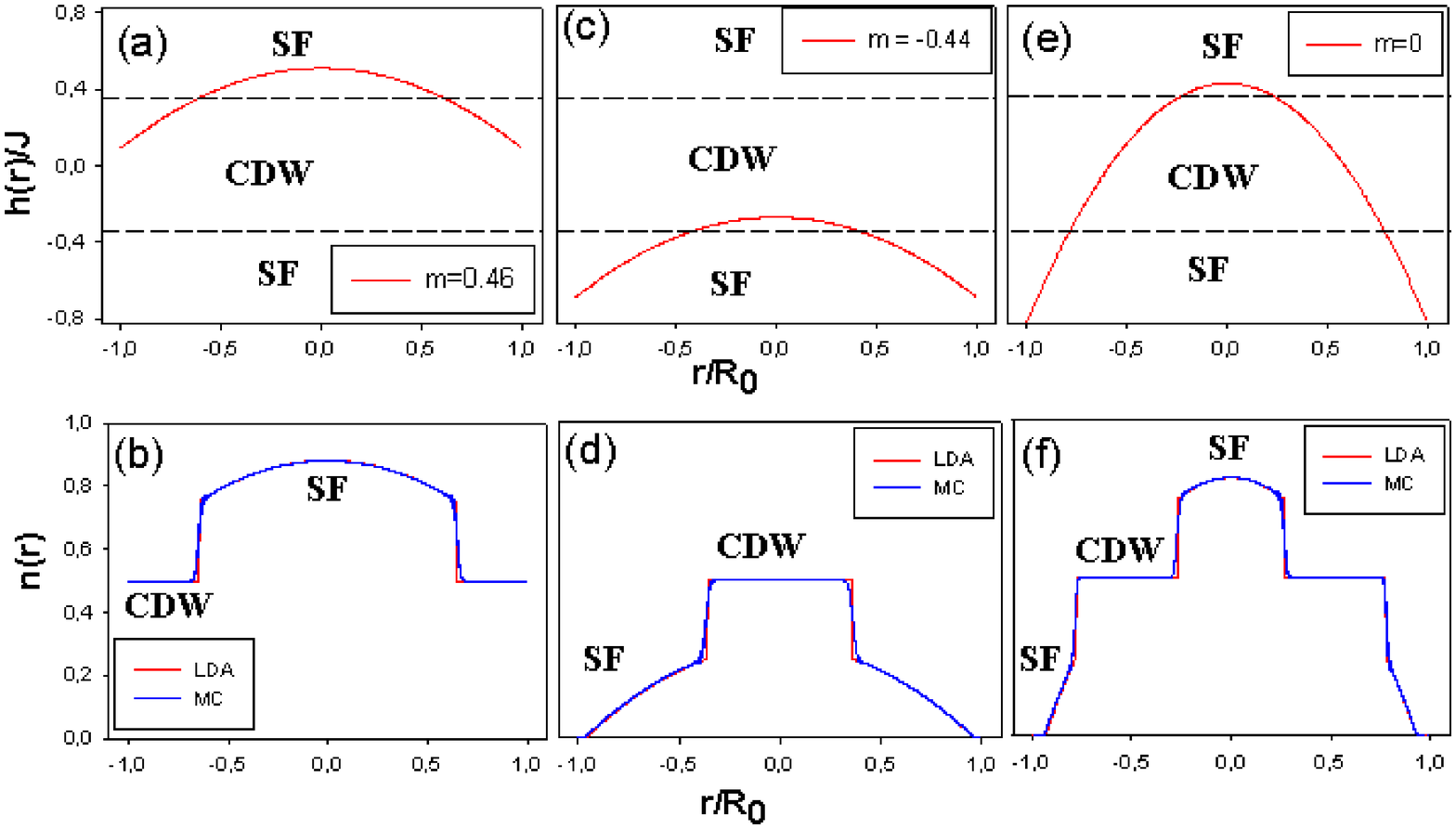}
  \caption{Density profiles and domains with different
  orderings inside the trap (bottom panels), as discussed
  in text. The top panels show how the trap potential
  intersects the characteristic values of the chemical potential
  in each case.
  }\label{Fig:1DXXZ}
\end{figure}

We finally discuss the effect of the trap potential. For
simplicity, we perform an explicit calculation only in the strong
coupling limit, using again the particle-hole transformation
(Table~\ref{table:ph}) and considering the effective spin model
(\ref{XXZ}). A harmonic trap potential yields a position-dependent
chemical potential which corresponds, under the particle-hole
transformation to a spatially varying magnetic field
$h(r)=h-h_{0}r^{2}/R_{0}^{2}$.
Here $R_0$ is the radius of the circular trap,
$h_{0}=m\omega^{2}_{o}R^{2}_{0}/2$ and $h=\mu-|U|/2$ is related to
the chemical potential at the center of the trap, which must be
adjusted so that the local density $n(r)$ integrates to the total
number of atoms. We start from a local density approximation
(LDA), and also compare with a Monte Carlo solution of the
strong-coupling model in the presence of $h(r)$.
As described above, the strong coupling analysis of the uniform
system yields a critical magnetic field (chemical potential) at
which $m(h)$ is discontinuous. For $\vert
h\vert<h_{c}=J\zeta\sqrt{\gamma(\gamma+2)}=
\frac{8z\zeta}{1-z^2}\frac{\t_\uparrow\t_\downarrow}{|U|}$, we
have a SDW$_{z}$ (CDW) phase, otherwise we have a SDW$_{xy}$ (SF)
phase.
Within the LDA approximation, this implies that in a region where
$\vert h(r) \vert$ is smaller (resp. larger) than $h_c$ we locally
observe SDW$_z$/CDW ordering (resp. SDW$_{xy}$/SF). According to
the values of the parameters $h$ and $h_0$, and noting that
$h-h_0<h(r)<h$, one finds several different regimes:

(i) $h -h_0 > h_c$ or $h<-h_c$. The trap potential is always
larger than $h_c$, or smaller than $-h_c$, so that the system is
in a SDW$_{xy}$ (SF) phase everywhere inside the trap, and the
density profile varies smoothly. (ii) $h > h_c$ and $\vert
h-h_0\vert < h_c$: in this case, $h(r)>h_c$ inside a circle of
radius $R_1=R_0\sqrt{(h-h_c)/h_0}$ centered at $r=0$. Hence, one
has phase separation into two distinct regions: SDW$_{xy}$(SF)
ordering within this circle, and SDW$_z$(CDW) in the outer ring
(Fig.~\ref{Fig:1DXXZ}, left panel). (iii) $h-h_0<-h_c$ and $\vert
h\vert< h_c$: we find again phase separation, with the opposite
spatial arrangement. The SDW$_{xy}$(SF) part is stable out or a
circle of radius $R_2=R_0\sqrt{(h+h_c)/h_0}$, inside which there
is a SDW$_z$(CDW) phase (Fig.~\ref{Fig:1DXXZ}, middle panel). (iv)
$h>h_c$ and $ h-h_0<-h_c$. Then, the magnetic field profile
crosses both $h_c$ and $-h_c$, so that there are three spatial
regions: $R<R_1$ where we find SDW$_{xy}$(SF), then the ring
$R_1<r<R_2$, where SDW$_z$(CDW) establishes, and finally an outer
ring $r>R_2$ with SDW$_{xy}$(SF) ordering (Fig.~\ref{Fig:1DXXZ},
right panel).

In the three last cases (ii-iv), in which phase separation occurs,
the LDA approximation predicts a jump of the magnetization at the
phase boundaries $R_1$ and $R_2$, corresponding to a jump of the
density in the original $U<0$ model (see also
\cite{lin_imbalance_shells_pra_2006}). In order to test this
prediction and assess the validity of LDA, we performed a
classical Monte Carlo simulation of model (\ref{XXZ}) in the
presence of a spatially dependent field $h(r)$. For simplicity,
this test was performed in a one-dimensional geometry. We find a
remarkable agreement between the LDA density profiles and the
Monte Carlo solution, which confirms that very sharp variations of
the local density indeed takes place at the boundary between
domains in cases (ii-iv).

In conclusion, for attractive interactions, the dominant effect of
the mass imbalance is to induce a competition between superfluid
and density-wave ordering. In the presence of an inhomogeneous
trap potential, both phases can be stabilized in different regions
of the trap, with rapid variations of the local density at the
phase boundaries. We note finally that, in the case of the
\Li6/\K40 mixture, a simple estimate shows that the mass imbalance
$z$ can be varied over a large range by changing the lattice depth
$V_0/E_R$ ($z\ll 1$ at small $V_0/E_R$ and $z\simeq 0.9$ for
$V_0/E_R\simeq 15$), so that the effects discussed in this work
may indeed be observable in this system.

\begin{acknowledgments}
We are grateful to
I.~Carusotto, F.~Chevy, P.~S.~Cornaglia, T.~Giamarchi, D.~Rohe, C.~Salomon and F.~Schreck
for useful discussions.
Support was provided by the ANR under contract "GASCOR", by
CNRS, Ecole Polytechnique and MIUR-PRIN Prot.200522492.
\end{acknowledgments}


\begin{thebibliography}{18}
\expandafter\ifx\csname natexlab\endcsname\relax\def\natexlab#1{#1}\fi
\expandafter\ifx\csname bibnamefont\endcsname\relax
  \def\bibnamefont#1{#1}\fi
\expandafter\ifx\csname bibfnamefont\endcsname\relax
  \def\bibfnamefont#1{#1}\fi
\expandafter\ifx\csname citenamefont\endcsname\relax
  \def\citenamefont#1{#1}\fi
\expandafter\ifx\csname url\endcsname\relax
  \def\url#1{\texttt{#1}}\fi
\expandafter\ifx\csname urlprefix\endcsname\relax\def\urlprefix{URL }\fi
\providecommand{\bibinfo}[2]{#2}
\providecommand{\eprint}[2][]{\url{#2}}

\bibitem[{\citenamefont{{Jaksch} and {Zoller}}(2005)}]{lattice_reviews}
\bibinfo{author}{\bibfnamefont{D.}~\bibnamefont{{Jaksch}}} \bibnamefont{and}
  \bibinfo{author}{\bibfnamefont{P.}~\bibnamefont{{Zoller}}},
  \bibinfo{journal}{Ann. Phys.} \textbf{\bibinfo{volume}{315}},
  \bibinfo{pages}{52} (\bibinfo{year}{2005});
\bibinfo{author}{\bibfnamefont{W.}~\bibnamefont{{Zwerger}}},
  \bibinfo{journal}{J. Optics B}
  \textbf{\bibinfo{volume}{5}}, \bibinfo{pages}{9} (\bibinfo{year}{2003});
\bibinfo{author}{\bibfnamefont{I.}~\bibnamefont{Bloch}},
  \bibinfo{journal}{Nature Physics} \textbf{\bibinfo{volume}{1}},
  \bibinfo{pages}{24} (\bibinfo{year}{2005});
\bibinfo{author}{\bibfnamefont{A.}~\bibnamefont{Georges}},
  \eprint{cond-mat/0702122}.

\bibitem[{\citenamefont{Cazalilla et~al.}(2005)\citenamefont{Cazalilla, Ho, and
  Giamarchi}}]{cazalilla_twomasses_prl_2005}
\bibinfo{author}{\bibfnamefont{M.~A.} \bibnamefont{Cazalilla {\it et al.}}},
  \bibinfo{journal}{Phys. Rev. Lett.} \textbf{\bibinfo{volume}{95}},
  \bibinfo{pages}{226402} (\bibinfo{year}{2005}).

\bibitem[{\citenamefont{{Werner} et~al.}(2005)\citenamefont{{Werner},
  {Parcollet}, {Georges}, and {Hassan}}}]{werner_cooling_2005}
\bibinfo{author}{\bibfnamefont{F.}~\bibnamefont{{Werner {\it et al.}}}},
  \bibinfo{journal}{Phys. Rev. Lett.}
  \textbf{\bibinfo{volume}{95}}, \bibinfo{pages}{056401}
  (\bibinfo{year}{2005}).

\bibitem[{\citenamefont{Duan}(2005)}]{duan_fermionhamiltonian_prl_2005}
\bibinfo{author}{\bibfnamefont{L.-M.} \bibnamefont{Duan}},
  \bibinfo{journal}{Phys. Rev. Lett.} \textbf{\bibinfo{volume}{95}},
  \bibinfo{pages}{243202} (\bibinfo{year}{2005}).

\bibitem[{inn()}]{innsbruck_KLiFR_2007}
\bibinfo{note}{Innsbruck group, private communication
  (http://www.uibk.ac.at/exphys/ultracold/)}.

\bibitem[{\citenamefont{Stan et~al.}(2004)\citenamefont{Stan, Zwierlein,
  Schunck, Raupach, and Ketterle}}]{stan_heteroresonance_prl_2004}
\bibinfo{author}{\bibfnamefont{C.~A.} \bibnamefont{Stan {\it et al.}}},
  \bibinfo{journal}{Phys. Rev. Lett.} \textbf{\bibinfo{volume}{93}},
  \bibinfo{pages}{143001} (\bibinfo{year}{2004}).

\bibitem[{\citenamefont{{Georges} et~al.}(1996)\citenamefont{{Georges},
  {Kotliar}, {Krauth}, and {Rozenberg}}}]{georges_review_dmft}
\bibinfo{author}{\bibfnamefont{A.}~\bibnamefont{{Georges {\it et al.}}}},
  \bibinfo{journal}{Rev. Mod. Phys.} \textbf{\bibinfo{volume}{68}},
  \bibinfo{pages}{13} (\bibinfo{year}{1996}).

\bibitem[{\citenamefont{Keller et~al.}(2001)\citenamefont{Keller, Metzner, and
  Schollw\"ock}}]{negUdmft}
\bibinfo{author}{\bibfnamefont{M.}~\bibnamefont{Keller {\it et al.}}},
  \bibinfo{journal}{Phys. Rev. Lett.} \textbf{\bibinfo{volume}{86}},
  \bibinfo{pages}{4612} (\bibinfo{year}{2001});
\bibinfo{author}{\bibfnamefont{M.}~\bibnamefont{Capone {\it et al.}}},
  \bibinfo{journal}{Phys. Rev. Lett.} \textbf{\bibinfo{volume}{88}},
  \bibinfo{pages}{126403} (\bibinfo{year}{2002});
%
\bibinfo{author}{\bibfnamefont{A.}~\bibnamefont{Toschi {\it et al.}}},
  \bibinfo{journal}{Phys. Rev. B} \textbf{\bibinfo{volume}{72}},
  \bibinfo{pages}{235118} (\bibinfo{year}{2005}{\natexlab{a}});
%
  \bibinfo{journal}{New. J. Phys} \textbf{\bibinfo{volume}{7}},
  \bibinfo{pages}{7} (\bibinfo{year}{2005}{\natexlab{b}});
%
\bibinfo{author}{\bibfnamefont{B.}~\bibnamefont{Kyung {\it et al.}}},
  \bibinfo{journal}{Phys. Rev. B} \textbf{\bibinfo{volume}{74}},
  \bibinfo{pages}{024501} (\bibinfo{year}{2006}).

\bibitem[{\citenamefont{Scalettar et~al.}(1995)\citenamefont{Scalettar,
  Batrouni, Kampf, and Zimanyi}}]{scalettar_bosehubbard_prb_1995}
\bibinfo{author}{\bibfnamefont{R.~T.} \bibnamefont{Scalettar {\it et al.}}},
  \bibinfo{journal}{Phys. Rev. B} \textbf{\bibinfo{volume}{51}},
  \bibinfo{pages}{8467} (\bibinfo{year}{1995}).

\bibitem[{\citenamefont{Freericks and
  Zlatic}(2003)}]{freericks_reviewFKmodel_rmp_2003}
\bibinfo{author}{\bibfnamefont{J.~K.} \bibnamefont{Freericks}}
  \bibnamefont{and} \bibinfo{author}{\bibfnamefont{V.}~\bibnamefont{Zlatic}},
  \bibinfo{journal}{Rev. Mod. Phys.} \textbf{\bibinfo{volume}{75}},
  \bibinfo{eid}{1333} (pages~\bibinfo{numpages}{50}) (\bibinfo{year}{2003}).

\bibitem[{\citenamefont{Lin et~al.}(2006)\citenamefont{Lin, Yi, and
  Duan}}]{lin_imbalance_shells_pra_2006}
\bibinfo{author}{\bibfnamefont{G.-D.} \bibnamefont{Lin {\it et al.}}},
  \bibinfo{journal}{Phys. Rev. A} \textbf{\bibinfo{volume}{74}},
  \bibinfo{pages}{031604} (\bibinfo{year}{2006}).

\end{thebibliography}


\end{document}